\documentclass[aps, prb, reprint, showpacs, showkeys, twocolumn, superscriptaddress, notitlepage]{revtex4-1}

\usepackage{float}
\usepackage{graphicx} 
\usepackage{braket}
\usepackage{color}
\usepackage[dvipsnames]{xcolor}
\usepackage{times}
\usepackage{textcomp}
\usepackage{amsmath}
\usepackage{lipsum}
\usepackage{multirow}  
\usepackage{adjustbox}  
\usepackage{textcomp} 

\DeclareTextSymbol{\degre}{T1}{6}
\DeclareTextSymbol{\degre}{OT1}{23}

\begin{document}

\title{Resonant driving of a single photon emitter embedded in a mechanical oscillator}
\vspace{1cm}

\author{Mathieu Munsch}
\author{Andreas V. Kuhlmann}
\author{Davide Cadeddu}
\affiliation{Department of Physics, University of Basel, Klingelbergstrasse 82, CH-4056 Basel, Switzerland}
\author{Jean-Michel G\'{e}rard}
\author{Julien Claudon}
\affiliation{Universit\'{e} Grenoble Alpes, F-38000 Grenoble, France}
\affiliation{CEA, INAC-PHELIQS, ``Nanophysique et semiconducteurs" Group, F-38000 Grenoble, France}
\author{Martino Poggio}
\author{Richard J. Warburton}
\affiliation{Department of Physics, University of Basel, Klingelbergstrasse 82, CH-4056 Basel, Switzerland}

\date{\today}

\begin{abstract}
Coupling a microscopic mechanical resonator to a nano-scale quantum system enables control of the mechanical resonator via the quantum system, and vice versa. The coupling is usually achieved through functionalization of the mechanical resonator but this results in additional mass and dissipation channels. An alternative is an intrinsic coupling based on strain. We employ here a monolithic semiconductor system. The nano-scale quantum system is a quantum dot; the mechanical resonator a microscopic “trumpet” which simultaneously optimizes the mechanical and photonic properties. The quantum dot transition is driven resonantly. Via the resonance fluorescence, we observe mechanical Brownian motion even at 4K, and demonstrate a coupling to mechanical modes of different types. We identify a mechanical mode with a cooperativity larger than one. We show analytically that the Heisenberg limit on displacement measurement can be reached with an embedded two-level system in the case of a transform-limited optical emitter with perfect photon detection. We argue that operation close to the Heisenberg limit is achievable with state-of-the-art quantum dot devices.
\end{abstract}

\maketitle

\section{Introduction}
Coupling a microscopic mechanical resonator to a nano-scale quantum system enables the mechanical resonator to be controlled via the quantum system, enabling ``phonon lasing"\cite{Kepesidis2013,Auffeves2014}, cooling towards the mechanical ground state\cite{WilsonRae2004,Kepesidis2013}, and generation of Schr\"odinger cat states\cite{Bassi2013}; and the quantum system to be controlled via the mechanical system, offering a route towards non-demolition read-out via a precise measurement of the oscillator's position\cite{Auffeves2014}. A futuristic application is to use the mechanical resonator as a bus to couple two remote qubits\cite{Rabl2010}

Mechanical resonators can be constructed on the micro- or nano-scale from various solid-state materials such as silicon nitride\cite{Thompson2008, Verbridge2008}, silica\cite{Armani2003}, silicon\cite{Eichenfield2009}, diamond\cite{Teissier2014}, GaAs\cite{Mahboob2008}, and so on. Position read-out of the mechanical resonator is usually carried out by incorporating the mechanical resonator in an optical cavity\cite{Aspelmeyer2014}. We pursue an alternative here, position read-out by embedding a single photon emitter into the mechanical resonator itself. 

GaAs is a natural choice of material for this endeavour. On the one hand, GaAs mechanical resonators are easy to make and have good mechanical properties\cite{Ding2010}. On the other hand, a self-assembled quantum dot in GaAs represents an excellent single photon emitter. At low temperature with resonant excitation, a quantum dot is a fast, bright\cite{Claudon2010,Somaschi2016} and pure\cite{He2013,Kuhlmann2015} source of single photons, outperforming any other solid-state emitter. A crucial feature is that the quantum dot transition frequency is sensitive to the strain produced by a deformation of the host material: there is an inherent coupling between the mechanics and the single photon emitter\cite{Yeo2014,Montinaro2014}. A recent insight is that semiconductor nano-wires have both mechanical resonances {\em and} an optimized photon collection efficiency\cite{Yeo2014}.   \\
\vspace{1.9cm}

\section{Quantum dots and opto-mechanics}

\begin{figure*}[t]
\includegraphics[width=0.92\textwidth]{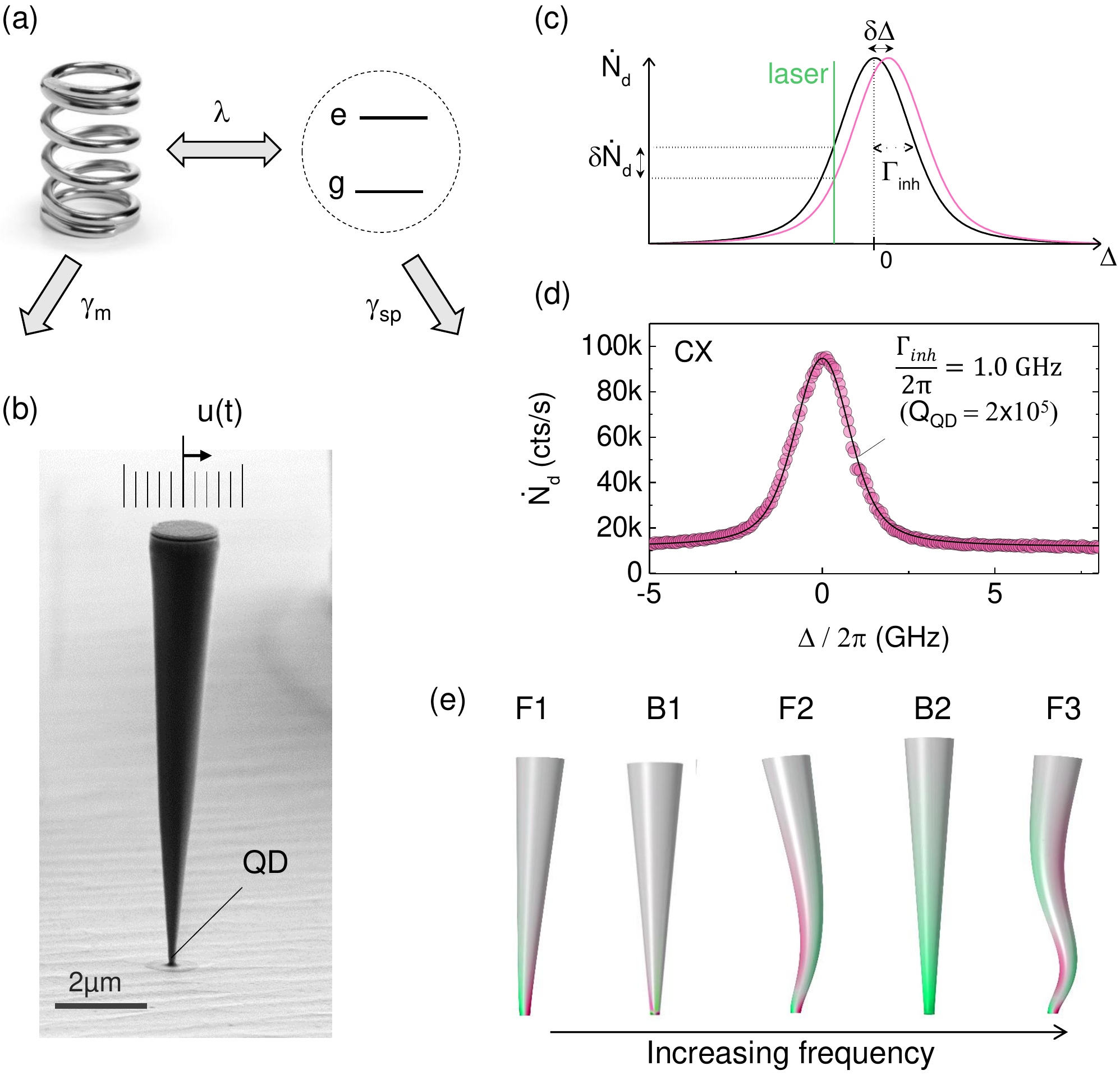}
\caption{{\bf The system}. (a) Sketch of the hybrid system: a mechanical oscillator is coupled to a two-level quantum system. The coupling rate $\lambda$ competes with the dissipation rate of the two parties: the intrinsic phonon relaxation rate $\gamma_m$ and the spontaneous emission rate of a photon $\gamma_{\rm sp}$. (b) Practical realization: a quantum dot (QD) is embedded close to the bottom of a micrometer-sized mechanical resonator. The coupling originates from inbuilt strain as the microwire oscillates. (c) A displacement $u$ of the microwire produces a shift $\delta \Delta$ in the QD frequency via intrinsic material strain, modifying the detuning between the QD and the laser. (d) Fluorescence signal from the charged exciton (CX) as a function of laser detuning (fluorescence wavelength $\sim 946$\,nm) . $\Omega_R \simeq \gamma_{\rm sp}= 1.1$ GHz. The fit uses a Voigt profile with a contribution to the linewidth of $0.45$ GHz from the Lorentzian part, and $0.70$  GHz from the Gaussian one. (e) The mechanical modes: F$_1$, F$_2$ and F$_3$ correspond to the first, second and third order flexural modes; B$_1$ and B$_2$ to two breathing modes. The colour map represents the strain within the trumpet (green: tensile, pink: compressive).}
\label{Figure1}
\end{figure*} 

The quantum dot (QD)--mechanical coupling manifests itself as a time-dependent frequency shift of the QD transition as the resonator oscillates, whose amplitude is determined by the hybrid opto-mechanical coupling $\lambda$\cite{Yeo2014,Montinaro2014} (Fig.~\ref{Figure1}a and c). This is described by the interaction Hamiltonian
\begin{equation}
\label{Hint2}
\hat{H}_{\rm int} = \frac{\hbar \lambda}{u_{\rm zpf}} \hat{u} \,\hat{\sigma}_z,
\end{equation}
where the Pauli operator $\hat{\sigma}_z$ acts on the QD two-level system, $\hat{u}$ is the operator representing the microwire's  displacement, and $u_{\rm zpf}$ corresponds to the quantum zero-point fluctuations. In previous experiments\cite{Yeo2014,Montinaro2014}, external driving of the mechanical resonator was used to demonstrate a coupling $\lambda$ of order 100 kHz, much less than the spontaneous emission rate, of order 1 GHz. A high quality QD inside such a resonator can thus be used to perform a sensitive read-out of the mechanical displacement. However, previous experiments employed non-resonant excitation which produces photons of low quality (as inferred from a large spectral linewidth and low indistinguishability) such that the displacement read-out via the QD photoluminescence would have been subject to a large amount of noise. 

We report here resonant optical driving of a high quality QD embedded in a mechanical system. The mechanical system, a semiconductor nanowire with inverse taper, a ``photonic trumpet" \cite{Munsch2013}, optimizes both mechanical and optical properties simultaneously, Fig.\ \ref{Figure1}b. The extra mass located at the end facet of the nanowire produces large strains in the ``stem" of the nanowire where the QD is located, resulting in large mechanical-optical couplings, Fig.\ \ref{Figure1}e. The QD emits photons preferentially into the one-dimensional waveguide mode defined by the nanowire. The mode then expands adiabatically by the inverse taper, allowing very large couplings to a Gaussian mode in free-space, resulting in very large photon extraction efficiencies. We exploit both features to detect thermal excitation of the mechanical resonances even at 4 K by measuring the fluctuations in the single photon count rate. The results open the way to adding and removing phonons to the mechanical oscillator via resonant optical driving of the QD. The results demonstrate a new approach to displacement read-out: the use of an embedded single photon emitter. In fact we show that the Heisenberg limit can be achieved in the limit of a perfect emitter (transform-limited linewidth, weak driving, 100\% collection efficiency), a regime within reach experimentally. 

\section{Detection of mechanical thermal motion}
To read-out the displacement of the mechanical resonator, we drive the optical transition of an embedded QD with a narrow band laser and collect the scattered light, the resonance fluorescence (RF). A displacement $u$ of the mechanical oscillator results in a detuning of the QD with respect to the constant frequency laser and translates into a change $\delta \dot{N}_d$ in the detected RF count rate. Assuming small optical detunings due to the mechanical oscillation,
\begin{equation}
\label{Nd_fluctuations}
 \delta \dot{N}_d = \frac{\alpha \, \hbar \lambda}{u_{\rm zpf}}\, u,
\end{equation}
where $\alpha = \partial \dot{N}_d/ \partial \Delta$ depends on the spectral profile of the emitter. In addition to the large opto-mechanical couplings and high photon extraction efficiency, the photonic trumpet has a further advantage: the large top facet serves as a ``landing pad" for the excitation beam and allows for excellent suppression of the back-scattered light from the resonant laser. 

We focus on a $12$ \textmu m long photonic trumpet, featuring a bottom diameter of $300$ nm and a top facet diameter of $1.62$ \textmu m.  The wire is clamped to a bottom gold-silica mirror via a flip-chip procedure. The typical spectrum from a QD inside such a wire is shown in Fig.\ \ref{Figure1}d. It is obtained from an excitation with two lasers: a very weak non-resonant laser is used to stabilize the QD's charge environment\cite{Nguyen2012} while a second laser scans the QD transition. We observe a maximum as the second laser hits the QD resonance ($\Delta =0$), on top of a photoluminescence background associated with the non-resonant pump. Thanks to the large top facet, the back-scattered resonant laser light is reduced by $40$ dB, which results in a negligible background (signal-to-background ratio is ${\rm S:B}\,=\,125$). In order to reach the best sensitivity of the QD to the mechanical motion, we have to maximize the count rate while maintaining a small linewidth. For this we operate at the onset of power broadening (Rabi coupling $\Omega_R \simeq \gamma_{\rm sp} $, with $\gamma_{\rm sp} = 1.1$ GHz, the spontaneous emission rate). This results in a linewidth $2\Gamma_{\rm inh} /2\pi = 2.0$\,GHz (Fig.\ \ref{Figure1}b) corresponding to an ``optical quality factor" $Q = 2 \times 10^{5}$.

\begin{figure}[t]
\includegraphics[width=0.48\textwidth]{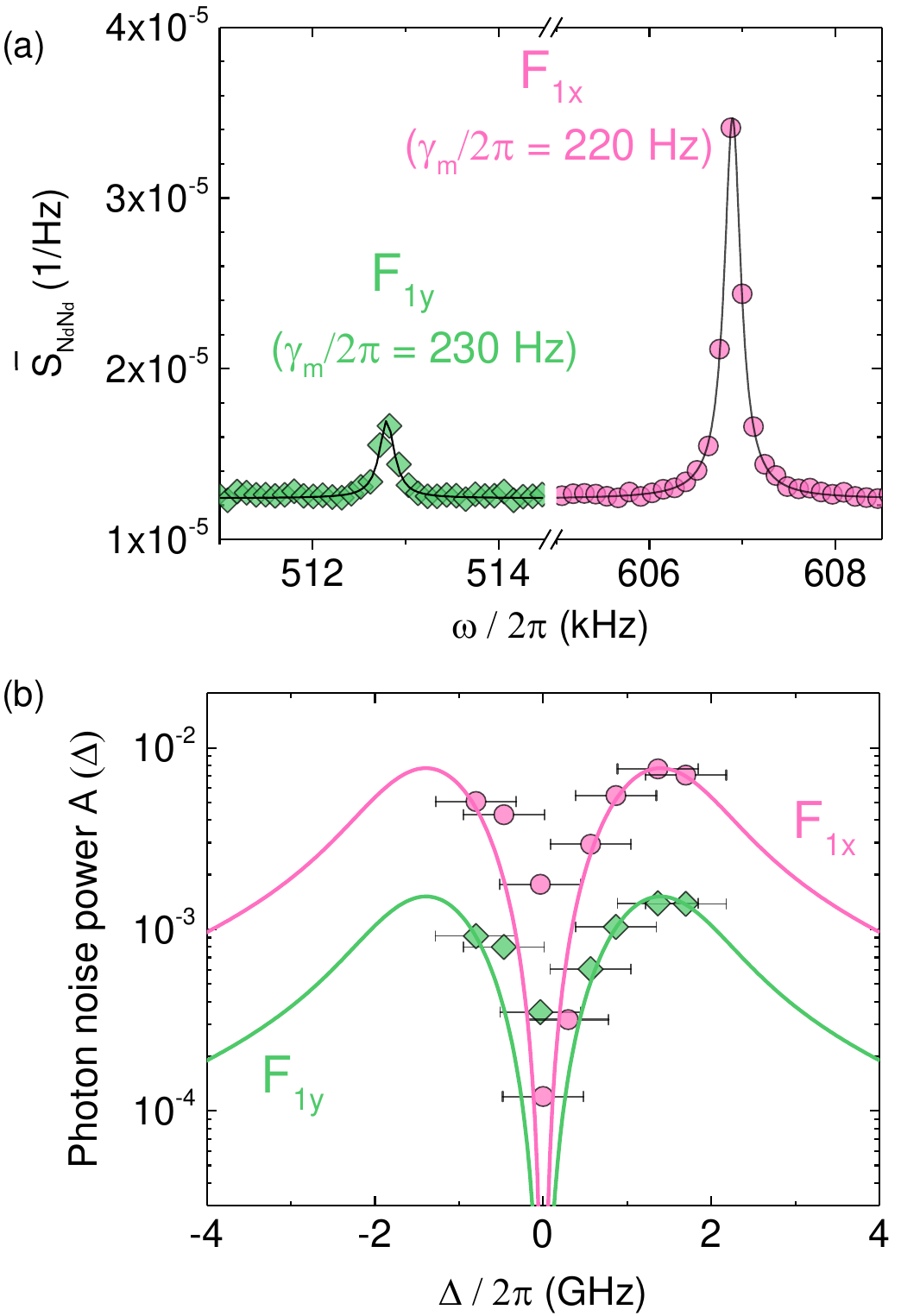}
\caption{{\bf Optical coupling to the fundamental mechanical mode}. (a) Photon noise power spectral density (NPSD) recorded at detuning $\Delta = \Gamma_{\rm inh}$ (integration time 20 min). The resonances correspond to Brownian motion of the first order flexural modes. The background corresponds to the shot noise level. (b) Photon noise power $\mathcal{A}$ as a function of laser detuning for F$_{\rm 1x}$ (pink circles) and F$_{\rm 1y}$ (green diamonds). The solid lines are fits to the experimental data with $\lambda$ as the only free parameter.}
\label{Figure2}
\end{figure}

We now turn to the noise spectroscopy of the RF signal and determination of the opto-mechanical coupling. One measurement protocol is to record a time-trace of the QD RF at a fixed laser detuning $\Delta$, Fig.\ \ref{Figure1}c, and perform a Fourier analysis on the data\cite{Kuhlmann2013a}. Fig.\ \ref{Figure2}a shows $\bar{S}_{\rm N_dN_d}$, the normalized noise power spectral density (NPSD), computed from a $20$ minute time-trace recorded at $\Delta = \Gamma_{\rm inh}$. The spectrum reveals two sharp resonances, labeled F$_{\rm 1y}$ and F$_{\rm 1x}$, at $\omega_{m, \rm F_{\rm 1y}}  /2\pi = 512.8$\,kHz and $\omega_{m, \rm F_{\rm 1x}} /2\pi = 607.9$\,kHz, respectively. These resonances, which are absent in the bulk sample, correspond to the thermally-driven mechanical resonances, i.e.\  the Brownian motion at $4$ K.  Specifically, we observe the two first order flexural modes, whose degeneracy is lifted by a slightly anisotropic cross section. The sensitivity of the measurement to the laser detuning is shown in Fig.\ \ref{Figure2}b, where we plot the noise power associated with each mode, i.e.\ the area $\mathcal{A} = \int \bar{S}_{\rm N_dN_d}(f)\,df$ under each peak. We get from Eq.\ \ref{Nd_fluctuations}
\begin{equation}
\mathcal{A}(\Delta) = \left(  \lambda\, \frac{u_{\rm th}}{u_{\rm zpf}} \frac{\alpha(\Delta) }{\langle  \dot{N_d}(\Delta) \rangle} \right)^2,
\label{Area}
\end{equation}
where $u_{\rm th}$ and $u_{\rm zpf}$ correspond respectively to the thermal and zero-point fluctuations. These quantities depend in particular on the characteristics of the mechanical resonator, namely its mode frequency and motional mass. The former is obtained from our measurement while the latter is determined through a finite element simulation of the resonator. $u_{\rm th}$ depends on the mode temperature, taken as 4 K assuming thermalization of the oscillator to the He bath. For the first flexural mode we find $u_{\rm zpf} = 2.3\,\times 10^{-14}$\,m and $u_{\rm th}\,=\,1.2\,\times 10^{-11}$\,m.
The detuning dependence of $\langle \dot{N_d}(\Delta) \rangle$ and $\alpha(\Delta)$ are obtained from a fit of the RF spectrum in Fig.\ \ref{Figure1}c so that Eq.\ \ref{Area} eventually only depends on the opto-mechanical coupling $\lambda$. 
Using $\lambda_{\rm F_{\rm 1x}}/2\pi = 280$ kHz and $\lambda_{\rm F_{\rm 1y}}/2\pi = 55$\,kHz, we find good agreement with the experimental data. 

\begin{figure}[tb]
\includegraphics[width=0.48\textwidth]{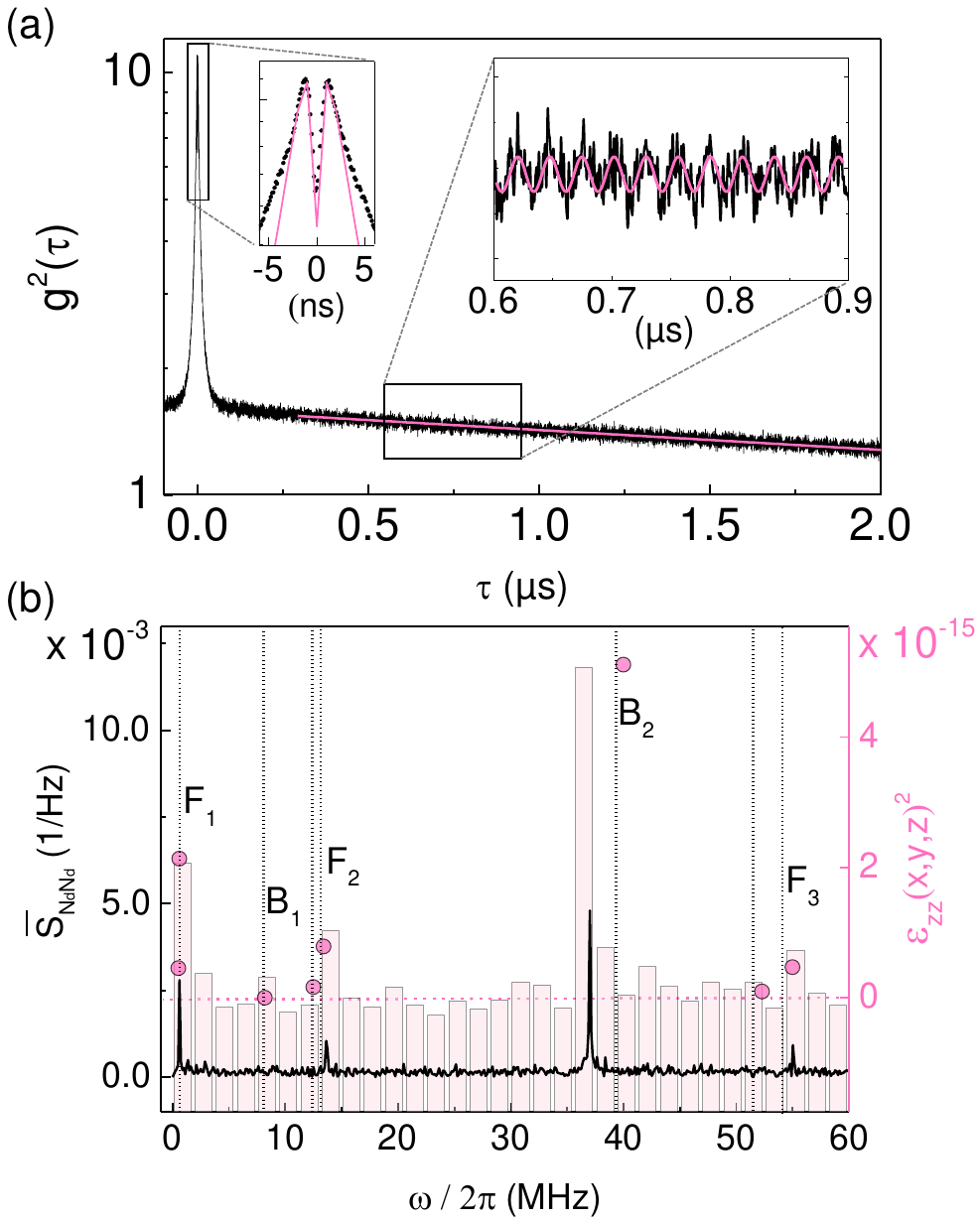}
\caption{{\bf Opto-mechanics in the photon counting regime}. (a) Auto-correlation function recorded at $\Delta \simeq \Gamma_{\rm inh}$ (the entire data extends over $8$ \textmu s). The oscillation in the right inset results from the mechanically induced modulation in count rate. The fit in the left inset (pink solid line) corresponds to a perfect single photon emitter subject to blinking ($\Omega_R = \gamma_{\rm sp}$) and includes the timing jitter of the detectors ($500$ ps). (b) In black, the NPSD obtained from the Fourier transform of the auto-correlation measurement. The black dotted lines correspond to the resonance frequencies obtained from a simulation. The bars in the background, corresponding to a 30 points binning of the same spectrum, are used for QD positioning. Right axis: the pink circles correspond to the expected relative amplitude of the modes ($S_{\dot{N}_d\dot{N}_d} \propto \epsilon^2$) obtained for a QD located $35$\,nm off axis with an angle $\phi = 20\degre$ from the $x$-axis. }
\label{Figure3}
\end{figure} 

The bandwidth of the previous measurement protocol is limited by the detector's dead-time. In our case, this means a cut-off at a frequency of $10$ MHz. To circumvent this limitation, we perform a correlation measurement of the RF signal with two detectors in a Hanbury Brown--Twiss configuration\cite{Matthiesen2014}. Fig.\ \ref{Figure3}a shows the result from a $70$ min measurement recorded at $\Delta = \Gamma_{\rm inh}$. The dip at zero delay is the signature of single photon emission from the QD. Its moderate depth is a consequence of the timing jitter of the detectors (note that the dip is narrowed by the unresolved Rabi oscillations induced by the amplitude of the drive, $\Omega_R \sim \gamma_{\rm sp}$). The bunching peak at short delays ($\tau < 100$\,ns) is related to a blinking in the QD emission\cite{Makhonin2014}. The maximum value of $g^{(2)}$ increases with the off:on ratio in the QD emission and the decay time of the bunching reveals the dynamics of charge fluctuations in the QD environment leading to the on:off behaviour\cite{Jahn2015}. In addition to these features, we observe a small oscillation, with a period of about $25$ ns, which runs over the entire 8 \textmu s time-span of the experiment. This oscillation corresponds to the signature of the opto-mechanical coupling in the photon counting regime. 

A Fourier analysis of the intensity correlation data at $\tau > 0.25$ \textmu s is shown in Fig.\ \ref{Figure3}. The spectrum reveals a whole series of resonances corresponding to different modes of the mechanical resonator. In particular, the lower frequency mode corresponds to the first order flexural mode (F$_1$) already evidenced in Fig.\ \ref{Figure2}, and the pronounced peak at $37$\,MHz corresponds to the second order breathing mode (B$_2$) immediately visible in the time-dependent data. Complete mode assignment is obtained from numerical simulation. With fine adjustment of the trumpet's dimensions (see Supplementary Information) we are able to reproduce the spectrum within a maximum error of 7.5\% in the exact frequency (vertical dotted lines in Fig.\ \ref{Figure3}b). In particular, we find that an ellipticity in the base diameter of 20\% accounts for the splitting observed in Fig.\ \ref{Figure2}b, in good agreement with earlier work\cite{Yeo2014}. 

From the relative amplitude of the various modes, it is possible to estimate the location of the QD in the nanowire with respect to the nanowire axis\cite{DeAssis2016}. This is shown on Fig.\ \ref{Figure3}b, where we plot the strain induced by the Brownian motion of each mode for a QD $35$ nm away from the central axis (see Supplementary Information for details). 

Quite remarkably, we observe a pronounced amplitude for B$_2$ despite the smaller phonon population associated with this high frequency mode ($u_{\rm th}\,=\,1.4\,\times 10^{-13}$\,m). This is the consequence of the large strain field associated with this specific mode (see Supplementary Information). Quantitatively, this translates in a coupling $\lambda_{\rm B_2}/ 2\pi = 3.6$ MHz, much larger than the values obtained for the coupling to F$_1$. The associated dissipation rate of B$_2$ also significantly increases to reach $\gamma_{\rm m,B_2} / 2\pi= 0.14$ MHz.  \\

\section{Quantum dot opto-mechanics: implications}

The measured noise spectrum may be translated into an equivalent displacement noise spectrum. For F$_1$, this results in a sensitivity $2.6 \times 10^{-13}$\, m.Hz$^{-1/2}$ ($\sqrt{S_{uu}} =  6.5 \times 10^{-14}$\, m.Hz$^{-1/2}$ for B$_2$). Although the value does not compete with the ones obtained from the best cavity opto-mechanics experiments\cite{Aspelmeyer2014}, we can nevertheless measure the zero-point fluctuations in just $70$\,s using our embedded single photon emitter. 

In the present case, the QD decay rate $\gamma_{\rm sp}$ is much larger than the phonon decay rate $\gamma_m$ and the coupling $\lambda$. Thus, the QD follows adiabatically the state of the mechanical oscillator. Its adiabatic elimination\cite{Gardiner2004} results in an effective dissipation rate for the mechanical resonator $\Gamma_{\rm opt} = \lambda^2 / \gamma_{\rm sp}$, which corresponds to an optically mediated damping (or anti-damping). 

For mode F$_1$, $\Gamma_{\rm opt} = 490$\,Hz \,$> \gamma_m$ implying a more efficient optically mediated damping compared to the intrinsic phonon relaxation. The cooperativity of the coupled single photon emitter--phonon system is $\mathcal{C}= \lambda^2/\gamma_{\rm sp} \gamma_{\rm m} = 2.2$, i.e. greater than one. This makes F$_1$ a very interesting mode for quantum applications. Focusing on B$_2$ we find $\Gamma_{\rm opt} = 0.08$\,MHz, with a cooperativity of $\mathcal{C}= 0.6$. We point out that B$_2$ has a homogeneous strain profile in the QD plane, Fig.\ \ref{Figure1}e, such that an on-axis quantum dot position simultaneously maximizes both the opto-mechanical coupling and the photon collection efficiency.

To implement the ``phonon laser" proposal\cite{Auffeves2014}, only the mechanical damping is important: the number of coherent phonons created through the optical driving of the two-level system depends on $(\lambda / \gamma_m)^2$. For both F$_1$ and B$_2$, $\lambda \gg \gamma_m$. For F$_1$, with the present trumpet, the scheme predicts that the QD generates $6 \times 10^6$ phonons while there are $1.7 \times 10^5$ thermal phonons at $4$\,K. \\

\section{Theory of position read-out using an embedded single photon emitter}
In standard opto-mechanics experiments, the mechanical oscillator is part of an optical cavity, allowing position read-out by measuring the response of the cavity. In this scenario, it is always possible to increase the intra-cavity power to reduce the shot noise. However, above a certain threshold, a second contribution associated to the noise from the back-action starts to dominate. This implies the presence of a minimum added noise, which, in the case of perfect collection efficiency, is referred to as the standard quantum limit (SQL) (see Ref.\ \onlinecite{Clerk2010} for a review). Here we present an alternative scheme: position read-out via the optical response of an embedded two-level quantum emitter. We find that the embedded quantum emitter enables the Heisenberg limit to be reached provided, first, the collection is perfectly efficient ($\epsilon =1$) and, second, the emitter has a transform-limited linewidth, $\Gamma= \gamma_{\rm sp}/2$. 

\begin{figure}[b]
\includegraphics[width=0.48\textwidth]{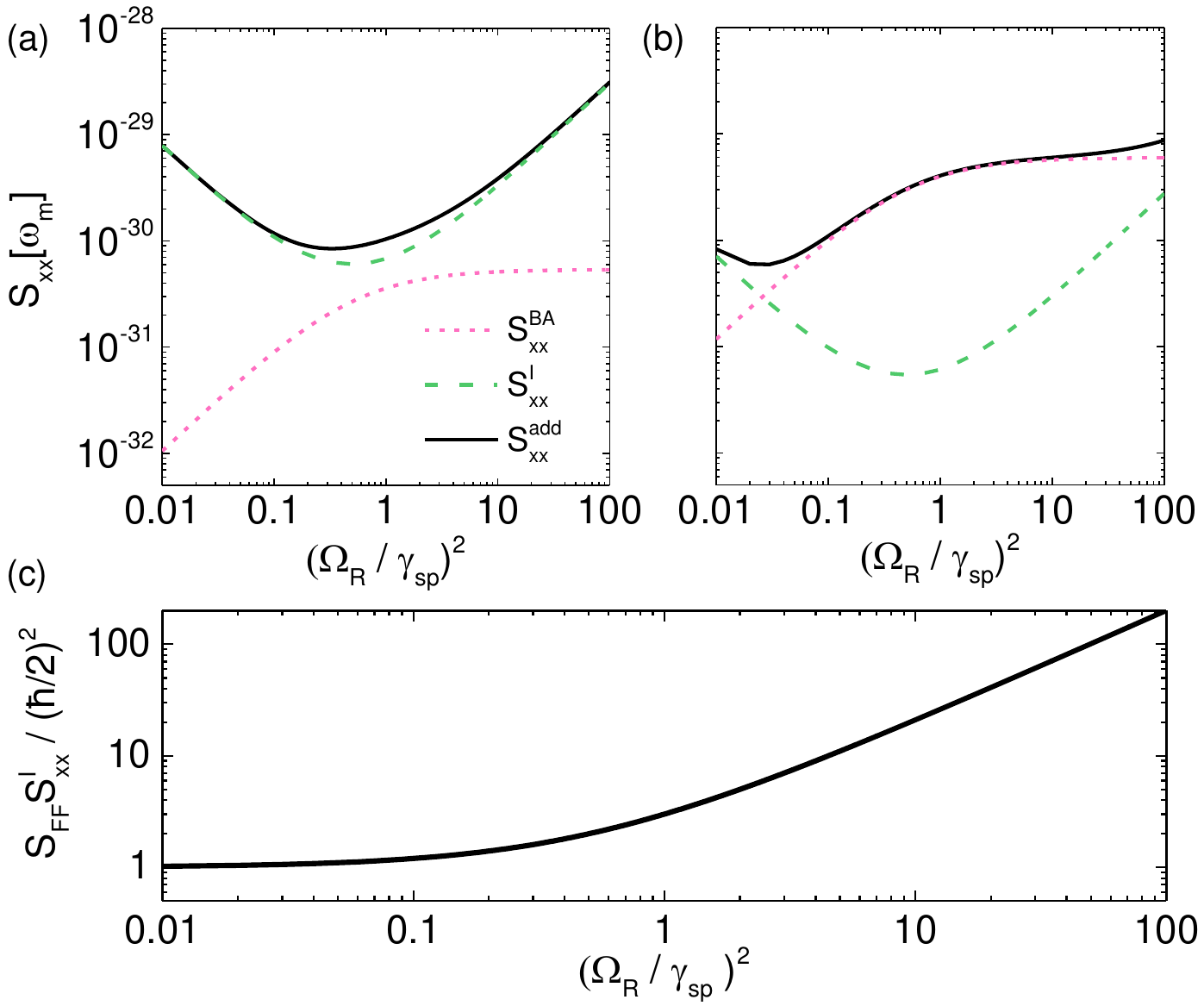}
\caption{{\bf Quantum position measurement with a qubit}. Evolution of the different contributions to the added noise $S_{xx}^{\rm add}$, as a function of the control parameter $\Omega_R$ for an ideal qubit probe ($\epsilon = 1$, $\Gamma = \gamma_{\rm sp}/2$). In (a) $\lambda^2 < \gamma_{\rm sp} \gamma_m$; in (b) $\lambda^2 \gg \gamma_{\rm sp} \gamma_m$. (c) Deviation from the Heisenberg limit due to optically induced power broadening of the transition ($\Gamma \propto \Omega_R$ in Eq.\ \ref{Heisenberglimit}).}
\label{Figure4}
\end{figure}

We sketch the proof of this result (see Supplementary Information for details). First, we focus on the imprecision noise resulting from shot noise on the collected photons. Using Eq.\ \ref{Nd_fluctuations}, the photon noise is translated into an equivalent displacement noise $S_{uu}^{I} = \left( \frac{u_{\rm zpf}}{\alpha\, \lambda} \right)^2  S_{\dot{N}_d\dot{N}_d}$, where $S_{\dot{N}_d\dot{N}_d}$ corresponds to the NPSD of the detected photons. Assuming no inhomogeneous broadening of the transition (the spectrum is Lorentzian, with a full width half maximum linewidth $\Gamma$), and operating at $\Delta = \Gamma$, $S_{uu}^{I}$ takes a simple form:
\begin{equation}
\label{SIxxp}
S_{uu}^{I} = \left( \frac{u_{\rm zpf}\, \Gamma}{ \lambda} \right)^2  \frac{1}{\dot{N}_d}.
\end{equation}
Fig.\ \ref{Figure4}a and b show the evolution of the imprecision noise as a function of the control parameter, the amplitude $\Omega_R$ of the resonant drive. As $\Omega_R$ increases, there is at first a reduction in the noise associated with the higher count rate and therefore a higher signal to noise ratio. However, in contrast to standard cavity-based opto-mechanics, we note the presence of a minimum after which the imprecision noise increases again. This is related to a coherent interaction between the two-level system and the optical driving field (Rabi oscillations), leading to power broadening ($\Gamma$ increases in Eq.\ \ref{SIxxp}) and a lower sensitivity. In addition to the imprecision noise, the read-out process eventually introduces noise due to the back-action exerted by the single photon emitter itself on the mechanical resonator. Here again, the situation diverges from standard cavity-based opto-mechanics since the system is now intrinsically non-linear. Solving the Heisenberg equation for the momentum operator, we find the back-action force, $\hat{F} =-\frac{\hbar \lambda}{u_{\rm zpf}} \hat{\sigma}_z$. 
For frequencies smaller than the spontaneous emission rate $\gamma_{\rm sp}$ the anti-bunched nature of the emission can be neglected and the back-action force noise reads
\begin{equation}
\label{SFFp}
S_{\it FF} = \left( \frac{\hbar \lambda}{u_{\rm zpf} \, \gamma_{\rm sp}}\right)^{2} \dot{N},
\end{equation}
where $N$ is the total number of emitted photons. The displacement NPSD associated to the back-action force is then simply obtained from $S_{xx}^{BA}[f]= |\chi_{xx} [f]|^{2} S_{FF} $, where $\chi_{xx} [f]$ is the mechanical susceptibility. 
This is shown in Fig.\ \ref{Figure4}a and b for $f = f_m$. $S_{xx}^{BA}$ increases initially as the two-level system is driven harder and then saturates as the population of the two-level system saturates. We find that the back-action dominates over the imprecision ($S_{xx}^{BA}[f_m] > S_{xx}^{I}$) when $\lambda^2\,>\,2\Gamma\,\gamma_m\,/\,\epsilon$ (see Supplementary Information) which, for $\Gamma = \gamma_{\rm sp}/2$ and using the optically mediated rate introduced earlier, reads $ \Gamma_{\rm opt} > \gamma_m/\epsilon$. Finally, by combining Eq.\ \ref{SIxxp} and \ref{SFFp} we find the central result for the product of the imprecision and back-action force
\begin{equation}
\label{Heisenberglimit}
S_{xx}^{I} \, S_{FF} \geq \frac{1}{\epsilon}\left( \frac{\hbar\,\Gamma}{\gamma_{\rm sp}} \right)^2.
\end{equation}
For perfect read-out, which means $\epsilon=1$ and $\Gamma= \gamma_{\rm sp}/2$, we recognize the Heisenberg principle, which sets the ultimate sensitivity of the measurement, the so-called standard quantum limit (SQL). 
Fig. \ref{Figure4}c shows how $S_{xx}^{I}S_{FF}$ increases above the SQL as $\Omega_R$ is increased.

In the present experiment, $\Gamma_{\rm opt} \simeq \gamma_m$. However, in particular because of the blinking of the QD fluorescence and the low quantum yield of the detector at these wavelengths, our overall detection efficiency only amounts to $\epsilon = 0.2$\%. In consequence, the dominant source of noise remains shot noise. However, there are strong reasons to believe that future experiments can operate close to the Heisenberg limit. First, quantum dots with close-to-ideal optical properties have been demonstrated in other experiments\cite{Kuhlmann2015}. Operation in the Coulomb regime eliminates the blinking; quantum dot linewidths match the transform limit on time-scales of $\sim 1$ \textmu s when the environment is ``frozen"\cite{Kuhlmann2013a}, and even on much longer time-scales in special circumstances \cite{Kuhlmann2015}. Second, use of optimized photonic trumpets, close-to-unity efficiency superconducting nanowire detectors\cite{Marsili2013} and a less lossy rejection of scattered laser light would all increase $\epsilon$: we estimate that $\epsilon \simeq 0.5$ is realistic. Achieving this quantum dot performance in such a mechanical oscillator would enable displacement measurement with a noise close to the Heisenberg limit and would compete with the very best cavity schemes; furthermore, the coupling opens possibilities to control the mechanical oscillator via the quantum dot and vice versa.

\section{Conclusions}
We have shown that a single quantum dot embedded in a mechanical resonator is a very sensitive probe of the mechanical motion. In particular, thermally-driven mechanical motion is revealed by fluctuations in the intensity of quantum dot resonance fluorescence. The measurement reveals a large hybrid coupling between the quantum dot and a whole series of mechanical modes. The relative strength of these interactions is used to determine the radial position of the QD within the resonator; modes with a particularly large coupling are identified. Presently, our position sensitivity is limited by an imperfect collection efficiency and charge noise in the device. We demonstrate that the Heisenberg limit applies for efficient collection of photons from a spectrally-pure emitter, a realistic proposition given that the two conditions have been met separately in other experiments. Present performance should be sufficient for driving of a coherent macrosopic motion by excitation of an embedded microscopic emitter\cite{Auffeves2014}. Slightly larger mechanical frequencies will facilitate operation in the resolved side-band regime, with the perspective of using an embedded quantum dot to cool the mechanical oscillator into the few-phonon quantum regime.\\

\noindent{\bf Acknowledgements}\\
We acknowledge financial support from NCCR QSIT and SNF project 200020\_156637. We thank N.\ L\"{o}rch, C.\ Bruder, J.\ Teissier and S.\ Starosielec for fruitful discussions; D.\ Riedel, D.\ Najer, J.-P.\ Jahn and J.\ Roch for assistance in the lab. Sample fabrication was carried out in the ``Plateforme Technologique Amont" and in CEA/LETI/MINATEC/DOPT clean rooms.\\

\appendix

\section{Theory} 

\subsection{Description of the hybrid coupling}
\label{description of the hybrid coupling}

Our device corresponds to a quantum two-level system, the QD excitonic transition, coupled to the motion of the mechanical oscillator via inbuilt strain. The interaction takes the general form of a longitudinal coupling described by 
\begin{equation}
\label{Hint1}
\hat{H}_{\rm int} = \frac{\hbar \lambda}{2} (\hat{b}^{\dagger}+\hat{b}) \hat{\sigma}_z,
\end{equation}
where $\lambda$ is the hybrid opto-mechanical coupling, $b^{\dagger}$ and $b$ are the phonon creation and annihilation operators, and $\sigma_z$ is the Pauli matrix operator acting on the 2-level system. Introducing $\hat{u}$, the operator associated with the resonator's displacement, the interaction Hamiltonian reads
\begin{equation}
\label{Hint2}
\hat{H}_{\rm int} = \frac{\hbar \lambda}{u_{\rm zpf}} \hat{u} \,\hat{\sigma}_z,
\end{equation}
where $u_{\rm zpf} = \sqrt{\frac{\hbar}{2 m \omega_m}}$ corresponds  to the zero-point fluctuations.
As a result of this parametric coupling, a displacement of the mechanical resonator results in a shift of the QD transition frequency, which can be used to read out the displacement. Conversely, the QD dipole produces a back-action force, which may drive the resonator's motion and ultimately limits the detection resolution. 

Fluctuations of the resonator's position are analyzed through their power spectral density (PSD), which is defined in a similar fashion to its classical counter-part\cite{Clerk2010}
\begin{equation}
\label{NSD}
S_{xx}[\omega] = \int_{-\infty}^{+\infty}dt e^{i\omega t} \braket{\delta \hat{x}(t)\delta \hat{x}(0)}.
\end{equation}
To read-out the displacement of the mechanical resonator, we record fluctuations in the resonance fluorescence (RF) signal for a fixed laser frequency. By shifting the QD resonance frequency, a displacement of the mechanical oscillator results in a change in the detected fluorescence count-rate $\dot{N}_d$, see Fig.\ \ref{Figure1}c. Assuming small frequency shifts due to the mechanical oscillation,
\begin{equation}
\label{dxdNrelation}
 \delta \dot{N}_d =  \frac{ \partial \dot{N}_d}{ \partial \Delta} \,\delta \Delta = \frac{\alpha \, \lambda}{u_{\rm zpf}}\, u,
\end{equation}
where $\Delta = \omega_{\rm QD} - \omega_{\rm laser}$ is the QD-laser detuning and $\alpha = \partial \dot{N}_d/ \partial \Delta$ is the slope of the QD spectrum at the corresponding detuning. It is clear that the sensitivity of the detection depends on the strength of the opto-mechanical coupling, but also on the spectral profile of the QD. The former depends on the QD position inside the mechanical resonator, the latter is characterized by the QD linewidth, or equivalently the QD quality factor.

\subsection{Quantum measurement with a two-level system}

In addition to the thermal fluctuations $S_{xx}^{th}[\omega]$ of the mechanical resonator, the read-out process introduces an additional noise $S_{xx}^{\rm tot}[\omega]$ so that the measured quantity is
\begin{equation}
\label{Stot}
S_{xx}^{\rm tot}[\omega] = S_{xx}^{th}[\omega] + S_{xx}^{\rm add}[\omega].
\end{equation}
There are two contributions to the added noise. The first one is the imprecision noise and is due to noise in the detection chain. The second one comes from the fluctuations in the back-action force exerted by the QD electric dipole on the mechanical resonator. Neglecting any correlations between the two sources of noise,
\begin{equation}
S_{xx}^{\rm add}[\omega] = S_{xx}^{I}[\omega] + S_{xx}^{BA}[\omega].
\end{equation}
We emphasize that the first term corresponds to an \emph{equivalent} displacement noise associated with the photon noise on the read-out, and not to actual fluctuations of the oscillator.
In the following we derive  analytical expressions for both contributions.

\subsubsection{Back-action noise}

\begin{figure}[t]
\includegraphics[width=0.4\textwidth]{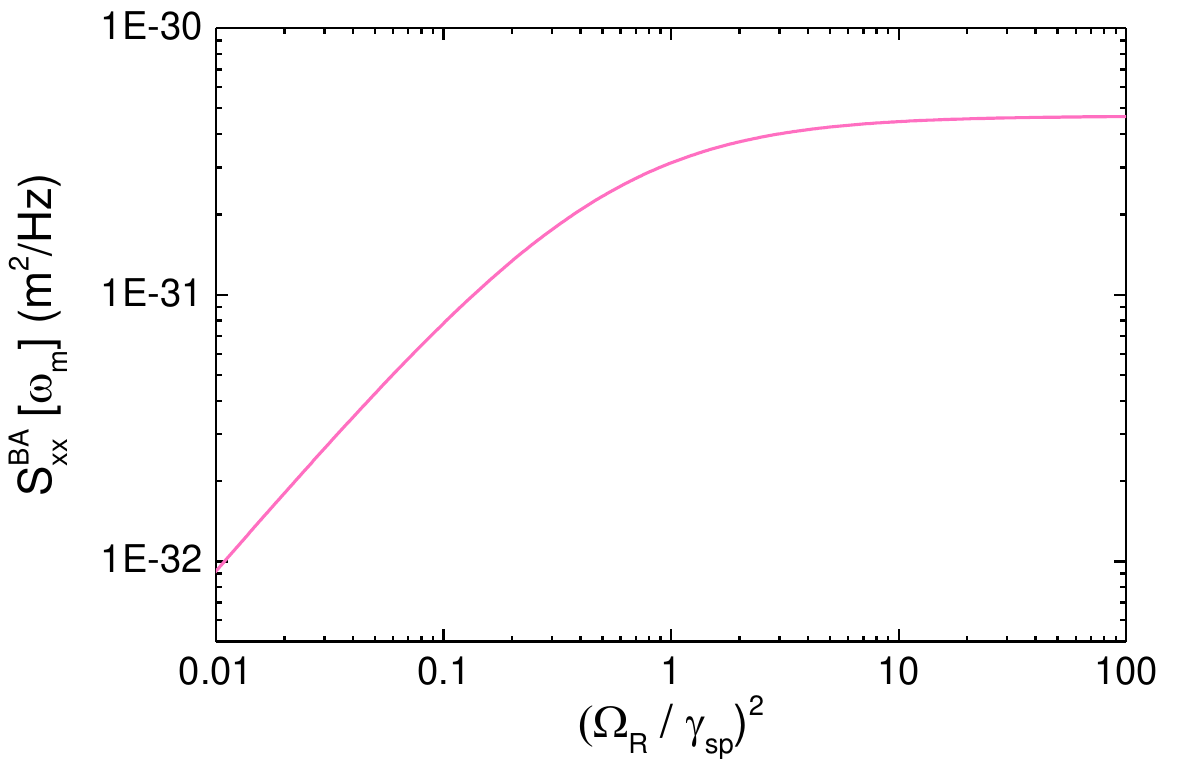}
\caption{Power spectral density of the back-action noise at $\omega=\omega_m$, in function of the optical drive $\Omega_R$. The mechanical oscillator has a mass $m_{\rm eff} = 2.6 \times 10^{-14}$\ kg. We use $\Gamma =\gamma_{\rm sp}/2$ with $\gamma_{\rm sp} = 1$\,GHz, $\gamma_m /2\pi= 300$ Hz and $\lambda /2\pi= 280$ kHz. }
\label{FigureS2}
\end{figure} 

The back-action force corresponds to the stress produced by the electric dipole upon excitation of the QD. Its expression can be derived by solving the Heisenberg equation for the momentum operator $\hat{p}$. This gives
\begin{equation}
\label{FBA}
\hat{F} \equiv \dot{\hat{p}}=\frac{1}{i \hbar} [H_{\rm int},p]=-\frac{\hbar \lambda}{u_{\rm zpf}} \hat{\sigma}_z.
\end{equation}
In our experiment, the QD dipole oscillates at frequencies much higher than the mechanical frequency $\omega_m$, so that we may consider the average value $\bar{\sigma}_z$. In particular, the resonance fluorescence signal from the QD is given by
$\dot{N}=\gamma_{\rm sp} \bar{\sigma}_z$,
where $\gamma_{\rm sp}$ is the spontaneous emission rate. 
Introducing $\Omega_R$, the amplitude of the resonant drive, it is standard quantum optics to derive the following expression \cite{Loudon1983}
\begin{equation}
\dot{N} (\Delta, \Omega_R) =  \frac{\gamma_{\rm sp}}{2} \frac{ \frac{\gamma}{\gamma_{\rm sp}}\, \Omega_R^{2}}{\Delta^2+\Gamma^2},
\end{equation}
where  $\gamma = \frac{\gamma_{\rm sp}}{2}+\gamma^{*}$ is the total QD decoherence rate (with $\gamma^*$ the pure dephasing rate) and $\Gamma = \sqrt{\gamma^2 + \frac{\gamma}{\gamma_{\rm sp}} \Omega_R^2}$ corresponds to the power-broadened half-width half-maximum linewidth(cf Fig.\ \ref{Figure1}c). (NB: In the main paper, $\Gamma_{\rm inh}$ includes additional spectral fluctuations.)
Coming back to Eq.\ \ref{FBA}, we derive a relationship between the fluctuations in the back-action force and the  fluctuations in the resonance fluorescence signal
\begin{equation}
\delta F = -\frac{\hbar \lambda}{u_{\rm zpf} \, \gamma_{\rm sp}} \delta \dot{N},
\end{equation}
and the corresponding expression for the power spectral density \begin{equation}
S_{\it FF}[\omega]= \left( \frac{\hbar \lambda}{u_{\rm zpf} \, \gamma_{\rm sp}}\right)^{2} S_{\it \dot{N}\dot{N}}[\omega].
\end{equation}
The anti-bunched nature of the QD emission is only visible in the correlations at short times. For $\omega \ll \gamma_{\rm sp}$ the QD photons thus have a white noise spectrum, so that $S_{\it \dot{N}\dot{N}} = \dot{N}$. This leads to the following expression for the noise spectral density of the back-action force
\begin{equation}
\label{SFF}
S_{\it FF} = \left( \frac{\hbar \lambda}{u_{\rm zpf} \, \gamma_{\rm sp}}\right)^{2} \dot{N},
\end{equation}
where $\dot{N}$ depends on the $\Omega_R$ and $\Delta$.
Finally, using the mechanical susceptibility
 \begin{equation}
 \chi_{xx} [\omega] = \frac{1}{m_{\rm eff}}\frac{1}{\omega_m^{2}-\omega^{2}- i \gamma_m \omega},
 \end{equation}
with $m_{\rm eff}$ the motional mass of the mechanical resonator, we obtain the displacement noise spectrum associated with the back-action of the two-level system 
\begin{equation}
\label{SxxBA}
S_{xx}^{BA}[\omega]= |\chi_{xx} [\omega]|^{2} S_{FF}= |\chi_{xx} [\omega]|^{2} \left( \frac{\hbar \lambda}{u_{\rm zpf} \, \gamma_{\rm sp}}\right)^{2}   \dot{N}.
\end{equation}
From this equation, we find that the back-action noise increases as we drive the system harder.  However, in contrast to standard linear opto-mechanics,  a maximum corresponding to achieving saturation of the excited state ($\bar{\sigma}_z = 1/2$) appears at high power, see Fig.\ \ref{FigureS2}.

\subsubsection{Imprecision noise}

\begin{figure}[t]
\includegraphics[width=0.4\textwidth]{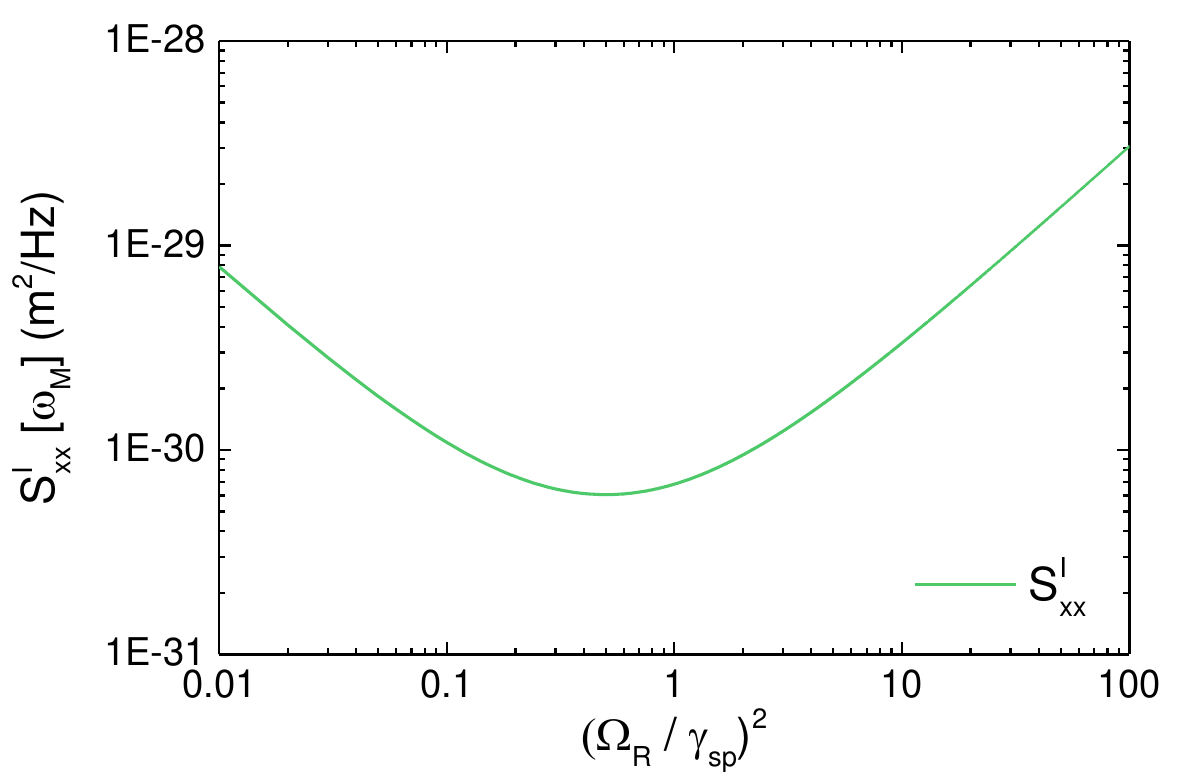}
\caption{Imprecision noise as a function of the optical drive $\Omega_R$. We use $\epsilon = 1$ and all other parameters are the same as for Fig.\ \ref{FigureS2}.}
\label{FigureS3}
\end{figure} 

\begin{figure*}[t]
\includegraphics[width=0.9\textwidth]{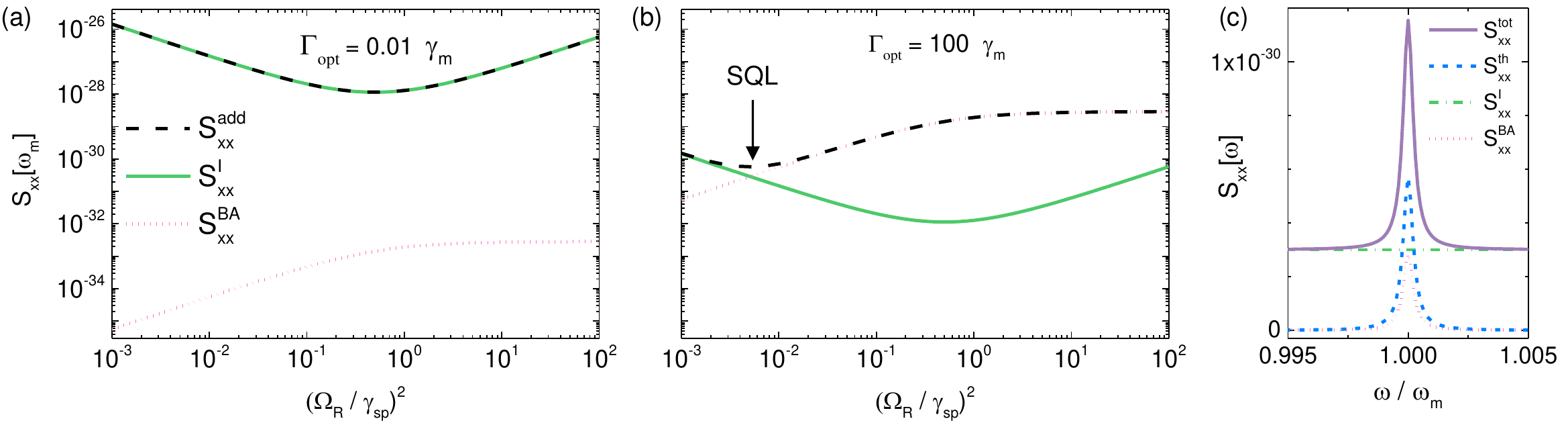}
\caption{(a) and (b) Power spectral densities of the \emph{added} noise and its two contributions. In (a) $\lambda = 0.1 \sqrt{\gamma_{\rm sp} \gamma_m}$, in (b) $\lambda = 10 \sqrt{\gamma_{\rm sp} \gamma_m}$. All other values are identical to those in Fig.\ \ref{FigureS3}. (c) Power spectrum of the \emph{total} noise (cf Eq.\ \ref{Stot}) and its different contributions. All parameters correspond to those of (b) and the optical drive is set to $\Omega_R/\gamma_{\rm sp} = 0.7$ (SQL). The temperature is set to $T=0$\,K, so that the thermal noise consists of just the quantum fluctuations.}
\label{FigureS4}
\end{figure*}

The imprecision noise corresponds to the uncertainty introduced by the measurement chain. Assuming perfect detectors, it boils down to shot noise on the detected count-rate (in practice the detectors are good enough so that this is a fair approximation). It originates from the Poissonian emission process and is amplified by the imperfect photon collection.
Introducing $\epsilon$, the overall collection efficiency, the detected count-rate simply reads $\dot{N}_d=\epsilon \dot{N}$.
Together with Eq.\ \ref{dxdNrelation}, it yields the following expression for the imprecision noise
\begin{equation}
\label{SIxx}
S_{xx}^{I} = \left( \frac{u_{\rm zpf}}{\alpha\, \lambda} \right)^2  S_{\dot{N}_d\dot{N}_d} = \left( \frac{u_{\rm zpf}}{ \lambda} \right)^2 \frac{\dot{N}}{\epsilon \,(\partial \dot{N}/ \partial \Delta)^2}.
\end{equation}
This equation takes a simple form for $\Delta = \Gamma$. At this point $\partial \dot{N}/ \partial \Delta = \dot{N} / \Gamma$, which results in
\begin{equation}
\label{SIxx2}
S_{xx}^{I} = \left( \frac{u_{\rm zpf}\, \Gamma}{ \lambda} \right)^2  \frac{1}{\epsilon \,\dot{N}}.
\end{equation}
As we increase the amplitude of the drive, we observe a decrease of the imprecision noise associated with the reduction of the shot noise, Fig.\ \ref{FigureS3}. Once again, the saturation of the two-level system imposes a limit to the minimum that can be reached. In fact, for $\Omega_R > \gamma_{\rm sp}$, $S_{xx}^{I}$ starts to increase as a result of the power broadening and the related lower sensitivity. Fig.\ \ref{FigureS3}. Finally, we note the presence of the detection efficiency in the denominator, which limits the ultimate achievable sensitivity.  

\subsubsection{Heisenberg principle and standard quantum limit in a hybrid system}

From equations \ref{SFF} and \ref{SIxx}, we can now write the product of the back-action force and the imprecision
\begin{equation}
S_{xx}^{I} \, S_{FF} = \frac{1}{\epsilon \,\gamma_{\rm sp}^2}\left( \frac{\dot{N}}{\partial \dot{N}/ \partial \Delta} \right)^2.
\end{equation}
This expression has a minimum for $\Delta = \Gamma$, which results in
\begin{equation}
S_{xx}^{I} \, S_{FF} \geq \frac{1}{\epsilon}\left( \frac{\hbar\,\Gamma}{\gamma_{\rm sp}} \right)^2.
\end{equation}
For perfect detection ($\epsilon = 1$), lifetime-limited QD and weak excitation ($\Gamma = \gamma_{\rm sp}/2$), we recognize the Heisenberg uncertainty principle
\begin{equation}
S_{xx}^{I} \, S_{FF} \geq \left( \frac{\hbar}{2} \right)^2.
\end{equation}
Equality is achieved for a laser detuning $\Delta = \Gamma$, and corresponds to the standard quantum limit (SQL). Interestingly, even for a lifetime-limited emitter, the SQL is only achieved in the weak excitation regime, see Fig 4 of the main paper. This is a consequence of the power broadening which reduces the sensitivity of the measurement for $\Omega_R \geq \gamma_{\rm sp}$. We emphasize that this is not the result of the complex solid-state environment around the QD, but is general to any driven two-level system coupled to a resonator through the interaction described in Eq.\,\ref{Hint1}. 

\subsubsection{Measuring the back-action}

Because of the inherent non-linearity of the two-level system, the contribution from the back-action may remain negligible and hard to detect.
To establish under which conditions the back-action dominates over the imprecision, we look for a solution to the equation
\begin{equation}
S_{xx}^{I}[\omega_m]-S_{xx}^{BA}[\omega_m]=0.
\end{equation}
This has an analytical solution at $\Delta = \Gamma$. Using Eqs. \ref{SxxBA} and \ref{SIxx2}, the problem reduces to solving a second order equation with respect to $\dot{N}$. After straightforward manipulation, we obtain for the positive solution
\begin{equation}
\dot{N}_+ = \frac{u_{\rm zpf}^2\,\Gamma\, \gamma_{\rm sp}}{\epsilon\, \hbar\, \lambda^2} \frac{1}{|\chi_{xx} [\omega_m]|}=\frac{1}{2}\frac{\Gamma\, \gamma_{\rm sp}\,\gamma_m}{\epsilon\, \lambda^2} .
\end{equation} 
Saturation of the resonance fluorescence imposes $\dot{N}_+ (\Delta = \Gamma) \leq \frac{\gamma_{\rm sp}}{4}$, which yields
\begin{equation}
\label{observeBA}
\lambda^2 \geq \frac{2 \Gamma\,\gamma_m}{\epsilon}.
\end{equation}
For $\epsilon = 1$ and assuming $\Gamma = \gamma_{\rm sp}/2$ (Fourier transform limit), we recognize the condition that the  optically mediated damping $\Gamma_{\rm opt} = \lambda ^2 / 2 \Gamma$ should be larger than the intrinsic phonon damping rate $\gamma_m$. In practice, the limited collection efficiency (finite quantum yield for the detectors, limited photon collection angle, blinking in the sample, etc) makes it more difficult to observe.

Fig.\,\ref{FigureS4} shows the two limiting cases. In (a), $\Gamma_{\rm opt} \ll \gamma_m$, in which case the minimum added noise corresponds to the minimum of $S_{xx}^{I}$, with a vanishing contribution from the back-action. In (b), $\Gamma_{\rm opt} \gg \gamma_m$, and the minimum, which corresponds to the SQL, is reached in the linear region. We then find a similar to standard cavity opto-mechanics experiments in the sense that, at the SQL, and for $\omega =\omega_m$, the back-action and the imprecision contribute equally to the noise, and the total added noise equals the zero-point fluctuations (Fig.\ \ref{FigureS4}c), i.e.
\begin{equation}
S_{xx}^{BA}[\omega_m] = S_{xx}^{I} = \frac{S_{xx}^{\rm th}[\omega_m,T=0]}{2} = \frac{ u_{\rm zpf}^2}{\gamma_m}.
\end{equation}

\section{Experiment}

\subsection{Resonance fluorescence}
The challenge with resonant spectroscopy is to distinguish between the fluorescence signal and the back-scattered laser light. For this we use a dark-field microscope based on cross-polarized excitation and detection. This technique ensures extinction ratios as high as $10^7$ upon reflection on a flat surface\cite{Kuhlmann2013b}. The situation is however more complex when the QD environment is processed below the micro-meter scale: a small object in the focus of the incident laser causes depolarization of the reflected beam and prevents efficient rejection.
To reconcile the nano-scale engineering of the QD surrounding with the need for a flat top facet, we use a tapered microwire, the ``photonic trumpet". A layer of QDs is embedded at the base of the structure where the lateral dimensions (between $200$ and $260$\,nm) forces the emission into the fundamental photonic mode propagating in the vertical direction\cite{Bleuse2011}. The subsequent tapered section results in an adiabatic deconfinement of the mode and limits diffraction losses at the top facet\cite{Munsch2013}. 
For the photonic trumpet studied in this paper (bottom diameter $300$ nm, top diameter $1.62$ \textmu m), we achieve a laser suppression $> 40$ dB over a $29$ GHz frequency span. This results in a signal to noise ratio $S:N = 125$ at a driving amplitude $\Omega = \gamma_{\rm sp}$.
To obtain a spectrum, we sweep the laser frequency with constant speed. The resulting resolution is 85 MHz, with an integration time of $0.2$\,sec/point.

\subsection{Detection efficiency}

At saturation, we detect a maximum resonance fluorescence count-rate of $0.83$\, MHz. Together with the measured $\gamma_{\rm sp} = 1.1$\,GHz, this yields $\epsilon = 0.16\,\%$ for the overall collection efficiency.
Taking the response of the detectors ($21 \%$), the transmission of our setup ($7 \%$) and blinking phenomena (QD ``on" time $10 \%$)  into account, we calculate the photon collection efficiency at the first lens (NA $= 0.8$) which amounts to $36\%$. This is about a factor two below previously reported measurement on similar structures\cite{Munsch2013}. We attribute the difference to the larger diameter at the QD position, which reduces the coupling to the fundamental guided mode, and to the position of the QD away from the axis (see Section.\ \ref{Quantum-dot localization}), a necessary condition to observe the response from the first order flexural mode (see also section \ref{Quantum-dot localization}).


\subsection{Evaluation of the coupling strength.}
\label{Evaluation of the coupling strength}

We determine the strength of the hybrid opto-mechanical coupling from the assumption that the oscillator is fully thermalized with the surrounding He bath (a realistic approximation since our sample is located in He exchange gas).  Using the equipartition theorem,
\begin{equation}
\langle  u^2 \rangle = \frac{k_B T}{m_{\rm eff} \omega_m^2}=u_{\rm th}^2.
\end{equation}
In addition, we have from Eq.\ \ref{dxdNrelation}
\begin{equation}
\langle  u^2 \rangle = \frac{u_{\rm zpf}^2}{\alpha^2\, \lambda^2}\, \langle  \delta \dot{N}_d^2 \rangle
\end{equation}
with $\alpha$ the derivative of the effective QD spectrum. Introducing the normalized photon noise spectrum $\bar{S}_{\rm \it NN}$,
\begin{equation}
\langle \delta \dot{N}_d^2 \rangle = \braket{\dot{N}_d}^2 \int\bar{S}_{\rm \it NN} (f)\,df,
\end{equation}
where $\int\bar{S}_{\rm \it NN} (f)\,df = \mathcal{A}$ is the photon noise power which is obtained for each mode from the area below the corresponding peak in the power noise spectrum (F1a or F1b, Fig.\ 2 of the main paper). This yields the following equation to describe the experimental data
\begin{equation}
\mathcal{A}(\Delta) = \left(  \lambda\, \frac{u_{\rm th}}{u_{\rm zpf}} \frac{\alpha(\Delta) }{\langle  \dot{N_d}(\Delta) \rangle} \right)^2,
\end{equation}
which only depends on $\lambda$. 

\subsection{Opto-mechanically induced dephasing}
\label{Optomechanically induced dephasing}

Throughout the analysis, it has been assumed that the exciton dephasing associated with the opto-mechanical coupling is negligible, such that it does not contribute to the broadening of the QD linewidth. This assumption, which allows here for an analytical theory to be developed (see Eq.\ \ref{dxdNrelation}), is also important in the context of quantum optics experiments with such devices. Indeed, the coupling to mechanical modes introduces additional noise, which leads to a broadening of the QD transitions for integration times longer than the mechanical oscillation period. 

The energy shift induced by the Brownian fluctuations of the microwire is simply given by
\begin{equation}
\delta \Delta_{\rm th} = \lambda\, \frac{u_{\rm th}}{u_{\rm zpf}}.
\end{equation}
We find $\delta \Delta_{\rm th, F_{1x}}/2\pi = 0.15$\,GHz and $\delta \Delta_{\rm th, B_2} /2\pi = 0.24$\,GHz for F$_{1x}$ and B$_2$ respectively, which correspond to the main sources of inhomogeneous broadening. This represent a total dephasing $\delta \Delta_{\rm th} \ll \Gamma_{\rm inh}$, which validates the hypothesis.

\begin{table*}[t]
\begin{ruledtabular}
\begin{tabular}{c|cccccc} 
\hspace{0.4cm} Mode\hspace{0.4cm} & $\omega_m$ (MHz) & $u_{\rm th}$ (pm) & $\epsilon_{\rm zz}$ & $\epsilon_{\rm xx}$ & $\epsilon_{\rm yy}$ \\
\hline 
$F_{1x}$  &  $0.6$ & $12$  & $5.9 \times 10^{-8}$ & $-1.6 \times 10^{-8}$ & $-1.9 \times 10^{-8}$\\
\hline
$B_{1}$  &  $8.2$ & $1.2$  & $1.3 \times 10^{-10}$ & $-2.6 \times 10^{-11}$ & $-1.2 \times 10^{-10}$\\
\hline
$F_{2x}$  &  $13.4$ & $0.5$  & $3.8 \times 10^{-8}$ & $-1.1 \times 10^{-8}$ & $-1.2 \times 10^{-8}$\\
\hline
$B_{2}$  &  $40.0$ & $0.1$  & $7.0 \times 10^{-8}$ & $-2.1 \times 10^{-8}$ & $-1.9 \times 10^{-8}$\\
\hline
$F_{3x}$  &  $55.0$ & $0.1$  & $3.0 \times 10^{-8}$ & $-8.1 \times 10^{-9}$ & $-9.7 \times 10^{-9}$\\
\end{tabular}
\caption{\label{strain} Simulated strain for a QD on the $x$-axis, 45 nm away from the centre (pink circle in Fig.\ \ref{FigureS8}a). Details of the geometry of the microwire are given in Section \ref{Numerical Analysis}.
}
\end{ruledtabular}
\end{table*}

\subsection{Numerical analysis}
\label{Numerical Analysis}

To confirm the origin of the resonances in our noise spectrum, we calculate the mechanical eigen-frequencies of the micro-resonator using a commercial finite-element analysis software (Comsol). We simulate a $12$ \textmu m long GaAs wire, with a bottom diameter of approximately $300$ nm and a tapering angle of $\theta = 3 \degre$. In order to adjust the mechanical frequencies to the experimental values, we allow a $5 \%$ variation on the length of the wire (due to flux inhomogeneities over the wafer surface in the MBE chamber). To account for the observed splitting of the first flexural mode, we also introduce a small asymmetry in the QD plane. In practice, the microwire has round top diameter but an elliptic base, the consequence of a slightly anisotropic etching process ($\Delta \theta = \theta_x -\theta_y$). 
Fig.\ \ref{FigureS8} shows the results for a $11.4$ \textmu m long wire, with a top diameter of $1.62$ \textmu m and a bottom section with minor axis $d_y=260$ nm and major axis $d_x =320 $ nm ($\Delta \theta = 0.15 \degre$\ , in good agreement with the estimation of the anisotropy in the etching process\cite{Yeo2014}).

\subsection{Quantum-dot localization}
\label{Quantum-dot localization}

\subsubsection{Complementary measurement}

\begin{figure}[b]
\includegraphics[width=0.48\textwidth]{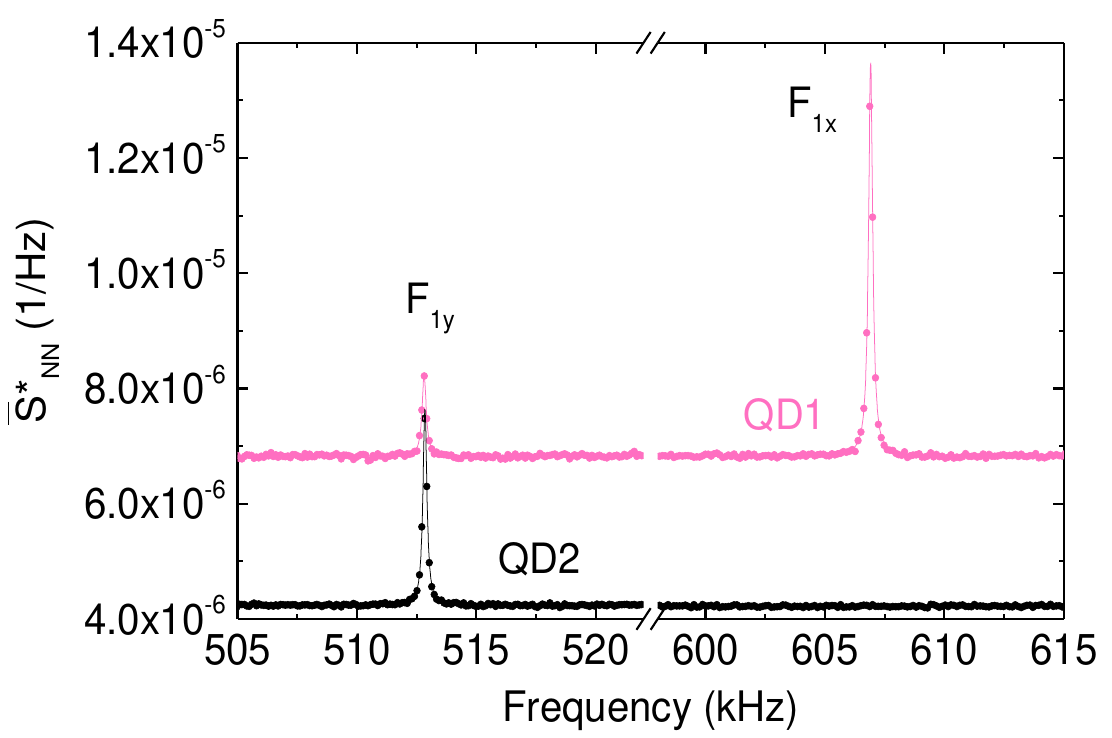}
\caption{QD noise spectrum for two different QDs (shot noise has been subtracted:  $\bar{S}_{\rm \it NN}^{*} =  \bar{S}_{\rm \it NN} - \bar{S}_{\rm \it NN}^{\rm shot}$). QD1 corresponds to the QD studied in the main paper. QD2 belongs to the same microwire and its noise spectrum is recorded in similar conditions: $\Omega_R \simeq \gamma_{\rm sp}$, $\Delta = \Gamma$, integration time of 20 mins.}
\label{QD2_data}
\end{figure} 

We present results from the noise spectroscopy on a second QD in the same microwire (QD2). The result is shown in Fig.\ \ref{QD2_data}, where we have removed the contribution from shot noise measured in a complementary measurement\cite{Kuhlmann2013a}. The different spectral signature is attributed to a different location of the QD in the microwire. In particular, the absence of a resonance at $607$ kHz means that the QD is located on the neutral (zero-strain) axis of F$_{1x}$. We point out that amplitude of the peak is not a direct measurement of the relative coupling strength, since the QD linewidths may be different. In the present case, QD2 has a larger FWHM (not shown)  which lowers the sensitivity of this second probe. The reduced noise floor is associated to this lower sensitivity, possibly combined with a less noisy charge environment around QD2. \\

\begin{figure*}[t]
\includegraphics[width=1\textwidth]{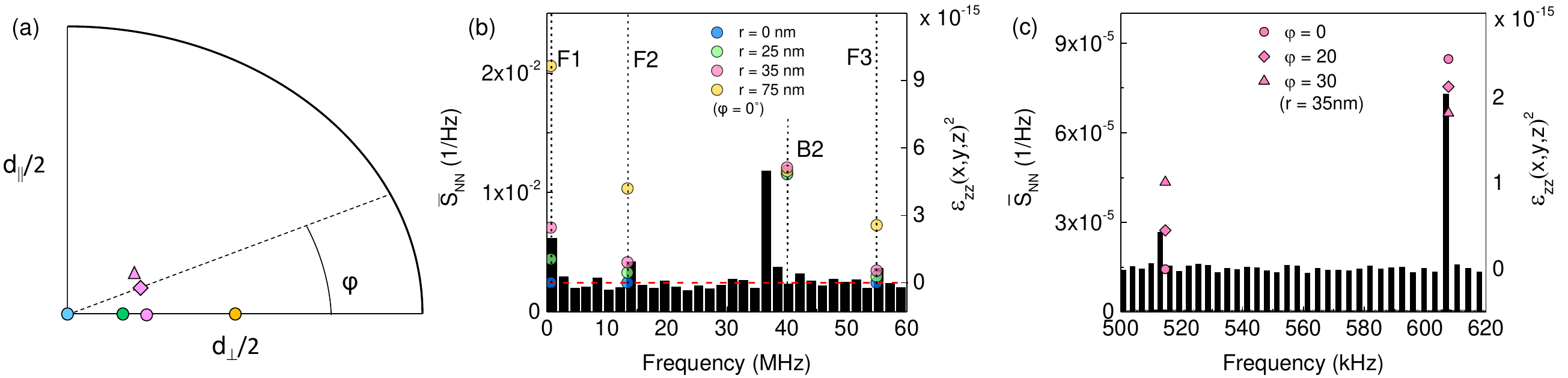}
\caption{QD localization. a) One quadrant of the QD plane. Coloured symbols correspond to various QD centre positions which result in the spectral signatures in b). The pink diamond corresponds to the position of QD1 (see Fig.\ 3 of the main paper). b) Expected relative amplitudes of the mechanical resonances for various QD positions along the perpendicular axis (coloured circles) compared to the experimental data (black bars). (Null values are not shown for clarity.) c) Effect of the azimuthal angle for a fixed distance to the centre.}
\label{FigureS8}
\end{figure*} 

\subsubsection{Full analysis}

A more precise analysis of the QD position within the trumpet is possible using the whole series of mechanical modes we have access to, see Fig.\ \ref{FigureS8}.
The idea is that each mode produces a specific strain at a given location in the wire. By comparing the relative amplitude of the measured resonances, it is possible to extract the QD position. In our case, the situation is simplified by the fact that we already know the precise location of the QD layer in the z-direction. 
The procedure goes as follows. We first simulate the strain per unit displacement for various radial positions of the QD at a fixed distance of $110$ nm from the base. For each mode, the data is normalized to the expected displacement associated with the Brownian motion. The result is shown in Table\ \ref{strain} for a QD located on the $x$-axis, at a $45$ nm distance from the center (pink circle in Fig.\ \ref{FigureS8}a). We observe that for all modes except B$_1$, the strain $\epsilon_{\rm zz}$ in the vertical direction dominates over the other components. In fact we find that $\epsilon_{\rm xx} \approx \epsilon_{\rm yy} \approx -\nu\,\epsilon_{\rm zz}$, where $\nu = 0.31$ is the Poisson ratio, meaning that we are dealing with uniaxial strain along the $z$-direction\cite{Stepanov2016,DeAssis2016}. Together with the same approximations as in Section \ref{description of the hybrid coupling} (linear interaction Hamiltonian, small frequency shifts), this results in $\bar{S}_{\rm \it NN} \propto \epsilon_{\rm zz}^2$.
We thus determine the QD position by adjusting position of the QD, calculating the relative amplitudes of the different modes and making a comparison with the result from the experimental data. 
This is illustrated in Fig.\ \ref{FigureS8} b for four different radial positions of the QD along the $x$-axis.
In a second step, we vary the angle to account for the different weights of the two orthogonal modes F$_{1x}$ and F$_{1y}$, Fig.\ \ref{FigureS8}c. 
Finally, using an iterative process, we are able to determine the QD position within the QD plane modulo a symmetry versus the $x$ and $y$ axes (this experiment does not resolve positive and negative coordinates). We find that QD1 is located $35$\ nm away from the axis of the trumpet, with an angle $\phi = 20 \degre$, see pink diamond in Fig.\ \ref{FigureS8}a and the corresponding fit in Fig.\ 3 of the main paper.

As a consistency check, we evaluate the QD frequency shift for the calculated strain. Neglecting confinement effects, this reads $\Delta^{\rm th}\,=\,a \epsilon_h + \frac{b}{2} \epsilon_{\rm sh}$, where $\epsilon_h = \epsilon_{xx} + \epsilon_{yy} +\epsilon_{zz}$ and $\epsilon_{\rm sh}= 2\epsilon_{zz} - \epsilon_{xx} - \epsilon_{yy}$ correspond to the hydrostatic and shear strains respectively, and $a$ and $b$ are material dependent deformation potentials\cite{Stepanov2016}. Assuming the QD is mainly composed of GaAs\cite{Munsch2014}, $a= -8.33$ eV and $b=-2.0$ eV\cite{Stepanov2016}, which gives $\Delta^{\rm th}_{F1x} = 0.08 $ GHz and $\Delta^{\rm th}_{B2} = 0.10 $ GHz. The results fall within the same order of magnitude as those obtained in Section \ref{Optomechanically induced dephasing}. 

\bibliographystyle{naturemag}







\end{document}